\documentclass{article}
\usepackage{xcolor}

\usepackage{arxiv}

\usepackage[utf8]{inputenc} % allow utf-8 input
\usepackage[T1]{fontenc}    % use 8-bit T1 fonts
\usepackage{hyperref}       % hyperlinks
\usepackage{url}            % simple URL typesetting
\usepackage{booktabs}       % professional-quality tables
\usepackage{amsfonts}       % blackboard math symbols
\usepackage{nicefrac}       % compact symbols for 1/2, etc.
\usepackage{microtype}      % microtypography
\usepackage{lipsum}
\usepackage{graphicx}
\graphicspath{ {./images/} }

\title{Using Reinforcement Learning to Allocate and Manage Service Function Chains in Cellular Networks}

\author{
 Guto Leoni Santos \\
  Centro de Informática\\
  Universidade Federal de Pernambuco\\
  Recife, Brazil \\
  \texttt{gls4@cin.ufpe.br} \\
   \And
  Patricia Takako Endo \\
  Universidade de Pernambuco \\
  Caruaru, Brazil \\
  \texttt{patricia.endo@upe.br} \\
}

\begin{document}
\maketitle

\begin{abstract}
It is expected that the next generation cellular networks provide a connected society with fully mobility to empower the socio-economic transformation. Several other technologies will benefits of this evolution, such as Internet of Things, smart cities, smart agriculture, vehicular networks, healthcare applications, and so on. Each of these scenarios presents specific requirements and demands different network configurations. To deal with this heterogeneity, virtualization technology is key technology. Indeed, the network function virtualization (NFV) paradigm provides flexibility for the network manager, allocating resources according to the demand, and reduces acquisition and operational costs. In addition, it is possible to specify an ordered set of network virtual functions (VNFs) for a given service, which is called as service function chain (SFC). However, besides the advantages from service virtualization, it is expected that network performance and availability do not be affected by its usage. In this paper, we propose the use of reinforcement learning to deploy a SFC of cellular network service and manage the VNFs operation. We consider that the SFC is deployed by the reinforcement learning agent considering a scenarios with distributed data centers, where the VNFs are deployed in virtual machines in commodity servers. The NFV management is related to create, delete, and restart the VNFs. The main purpose is to reduce the number of lost packets taking into account the energy consumption of the servers. We use the Proximal Policy Optimization (PPO) algorithm to implement the agent and preliminary results show that the agent is able to allocate the SFC and manage the VNFs, reducing the number of lost packets.
\end{abstract}

% keywords can be removed

\section{Introduction}
\label{sec:introduction}
The next generation of cellular mobile networks, 5G, is being developed to address several application requirements, such as massive number of heterogeneous devices, high transmission rate, low latency, and high reliability \cite{schulz2019network}. Internet of Things (IoT), smart city, vehicular networks, Industry 4.0, and other scenarios bring new types of mobile devices with heterogeneous hardware specification that will requires different network specifications. For example, the network requirements for a mobile gaming (while real time and fast connection are crucial) is different of a massive devices scenario, such as open air festival and sport events (where huge amount of resource and bandwidth are needed) \cite{kusume2015updated}. It is expected that the global Internet traffic will reach around 30 GB per capita by the 2021, and 63\% of this traffic is about mobile devices \cite{maksymyuk2018deep}. Therefore, the 5G network architecture need to be flexible and adaptable for different scenarios and applications.

Chen et al. \cite{chen2018reinforcement} divided the architecture of 5G network in three main components: physical structure layer, virtualization layer, and application layer. The physical layer is composed of different static equipment such as compute, storage, and network devices, that can be located in data centers. The virtualization layer is responsible to abstract the access to the physical layer through virtual resources pools with the purpose to make easy the resource sharing and utilization for different applications. The application layer provides the services to the end users through an interface provided by the virtualization layer.

This division of the network architecture allows operators to achieve a high level of flexibility, since resources can be allocated on demand according to the application traffic \cite{li2018deep}. In fact, the virtualization paradigm is a crucial key to meet different users requirements described previously. Unfortunately, network functions still depend largely on physical and dedicated devices (middleboxes or network appliances) \cite{bhamare2016survey}. To deal with this limitation, the network virtualization function (NFV) paradigm proposes to implement these functions in general purpose hardware (such as common servers) running in virtual machines \cite{moualla2018availability}. The interconnection of two or more virtual network functions (VNFs) for a complete end-to-end service is named service function chain (SFC). Thus, in a cellular network scenario, a SFC may be the core telecom stack \cite{bhamare2016survey}, and can be deployed in virtual machines located in distributed data centers. This new configuration increases the flexibility of the network and at the same time reduces operating expenditures (OPEX) and capital expenditures (CAPEX) \cite{herrera2016resource}.

However, the integration between cellular network and NFV to compose new 5G network architectures is not without challenges. It is expected that the performance and availability of applications are maintained with the use of NFV. As highlighted by Gupta et al. \cite{gupta2017colap}, maintaining the uptime of cellular networks is one of major issues from the operators' point of view. However, several aspects can impact the services hosted in data centers, such as software failures, misconfiguration, planned and unplanned downtime, and so on \cite{endo2017minimizing}.

The management of these networks is a complex task. Considering the network status before allocating an SFC can be a promissory way to ensure high level of quality of service \cite{xiao2019nfvdeep,wang2020sfc}. Additionally, a smart strategy need to be employed to guarantee the service high availability (for example, adding redundant VFN of a service) taking into account constraints, such as servers utilization and energy consumption. An approach that is able to learn according with different scenarios could be employ in this scenario, adapting the solution (SFC allocation) according to specific requirements and network conditions.

% allocating the SFC automatically and realizing changes on the environment is desirable for network managers.

In this paper, we use a reinforcement learning algorithm to allocate and manage SFC automatically in data centers considering a cellular network scenario. We create a scenario based on a data set of Milan region, Italy, that consists of internet activities from users related to 62 days \cite{barlacchi2015multi}. The scenario considered in this paper is composed of actual cells of the Milan city and distributed data centers to manage the traffic from users (by allocating SFC). The SFC based on services of the Evolved Packet Core (EPC), which is the core network of Long Term Evolution (LTE). 
% We consider that the scenario is dynamic, and the servers and VNFs executing in the data centers may fail eventually, impacting the SFC availability. 

A simulator was developed to represent the users' Internet activities based on the data set; and also to represent eventual failures and repair of the EPC network components. Moreover, we conduct experiments to create clusters of cells based on the Internet activities similarity. 
Then, the reinforcement learning agent can learn the best strategy to allocate VNFs in servers of distributed data centers, considering that both servers and VNFs may fail during their operation. While new Internet activities arrives in the data center, the agent can create, delete, and restart the VNFs in order to reduce the number of failures. The main purpose is to reduce the number of failures and, consequently, reduce the number of lost packets.
% Then, a reinforcement learning agent can learn how to handle the Internet activities of a specific cluster. 
An agent based on Proximal Policy Optimization (PPO) algorithm was trained in the environment and preliminary results show that it is able to allocate the SFC after some steps and manage it controlling the number of lost packets due failures.

\section{Material and Methods}
\label{sec:system-model}
\subsection{Scenario}
\begin{figure}[h]
\centering
\includegraphics[width=.5\textwidth]{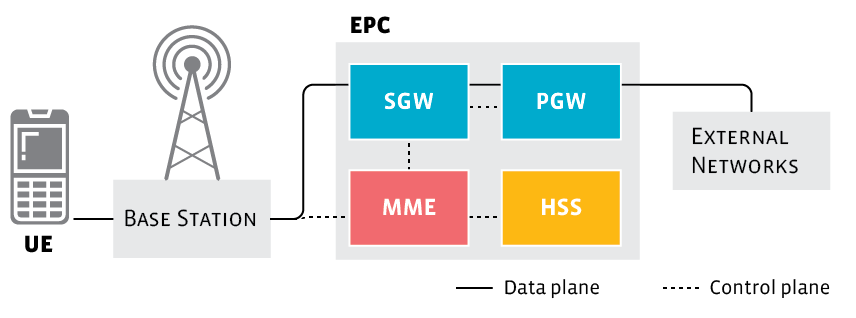}
\caption{EPC main functions}
\label{fig:epc-functions}
\end{figure}

We consider a set of distributed data centers located close to the base stations with VNFs related to the EPC core stack. The EPC is the core of network of LTE, where the control and data traffics are separated and addressed by different entities \cite{mohammadkhan2016cleang}, as shown in the Figure \ref{fig:epc-functions}. 

We consider that all components of EPC, the serving gateway (SGW), the packet gateway (PGW), the mobility management entity (MME), and the home subscribe server (HSS) will compose the SFC that will be deployed over the distributed data centers. Thus, each EPC component correspond to a VNF which can have redundant instances in order to increase the SFC availability.

We assume that all servers of a data center are connected among them, and all data centers are connected, i.e., any server has connection to all other operational servers in the data centers. We also consider that each server has a maximum limit of VNFs that can be deployed, representing the limited computation resource of such server. The both servers and VNFs run with a probability failure, which increases over time.

The servers of data centers can eventually fail for different reasons such as problems in storage, memory and processor failures, and so on. Due to these failures, the server and its respective virtual machines become unavailable. In a similar way, when a VNF is created, it can fail due to many reasons (software failures, bugs, unexpected maintenance, and so on). Both server and VNFs can be repaired after a failure, becoming operational again. We also consider that in the case of a server failure, the state of VFNs are maintained, i.e., after the repair, all VNFs are restored and can be process data from cellular users. We consider a real data set of cellular traffic as described in the next subsection.

\subsection{Data set}
\label{subsed:tim-dataset}
To represent the Internet traffic data, we used the Telecom Italia data set provided by \cite{barlacchi2015multi}. This data set provides a multi-source aggregation of telecommunications, news, social networks, and electricity data of two different regions of Italy: the Milan region and the Province of Tentrino.

\begin{figure}[h!]
\centering
\includegraphics[width=0.35\columnwidth]{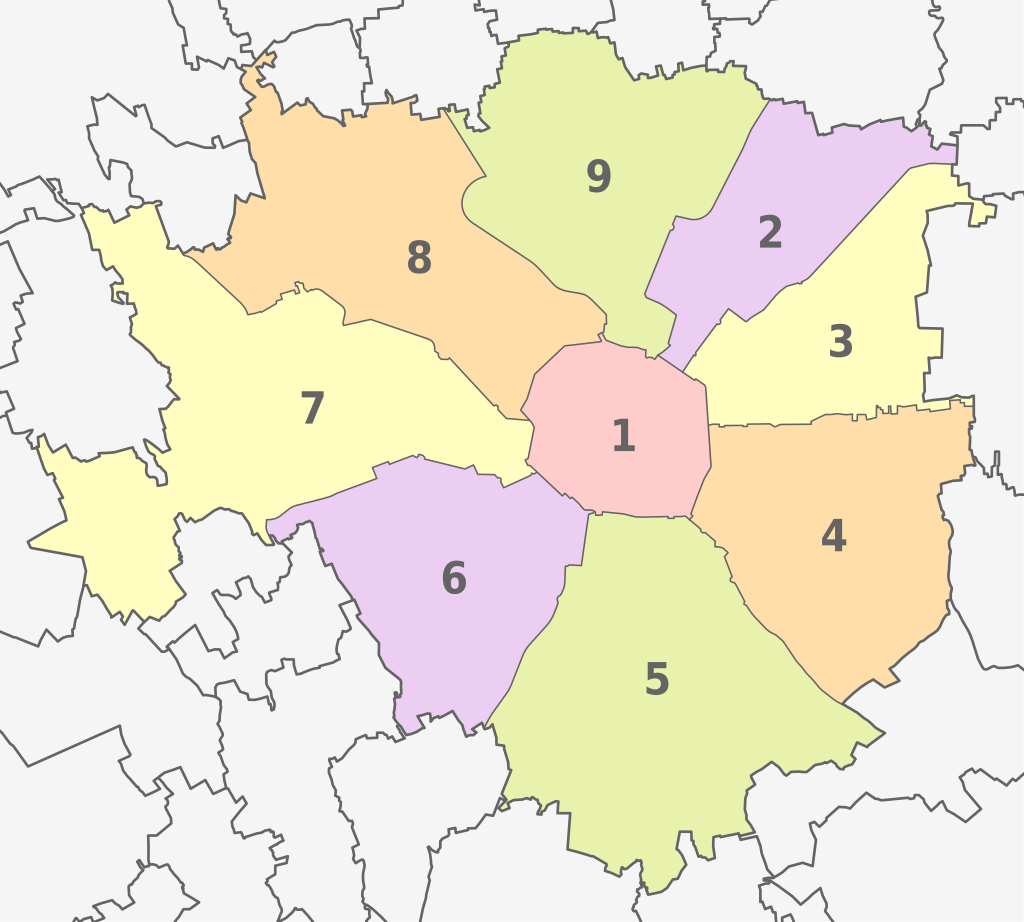}
\caption{The Milan Metropolitan Area. Source: \url{https://www.wikiwand.com/en/Municipalities\_of\_Milan}}
\label{fig:milan-region}
\end{figure}

From the data set, we use Internet activities data of the Milan region, composed of nine municipalities (see Figure \ref{fig:milan-region}), that is largest metropolitan area in Italy and one of most populous areas of the European Union \cite{eurostat20}. The data was collected between 1 November 2013 and 1 January 2014, comprising 62 days. 

The Milan region was divided into 10,000 cells (100 x 100) and contains about 10 million user activity logs related to a particular cell. The data set is composed of different Call Detail Records (CDRs), but in this paper, we focus only on Internet activities (call and SMS activities will be considered in future works). 

\subsection{Creating clusters}
Different regions of Milan may have different Internet activities patterns. For example, Internet activities in regions further from Milan's center may behave differently. Therefore, we analyze Internet traffic of cells in order to create clusters based on similarity.

We consider the average of Internet activities per day periods to represent the cells, and we used the K-means algorithm to create the clusters \cite{arora2016analysis}. For that, we define different periods of the day and calculate the number of Internet activities per period, as shown in the Table \ref{tab:periods-day}. Afterwards, we calculate the average Internet activity for each period for all cells present in the data set.

\begin{table}[h]
\caption{Periods of the day}
\label{tab:periods-day}
\centering
% \resizebox{\columnwidth}{!}{
\begin{tabular}{cc}
\hline
\textbf{Period} & \textbf{Time (in hour)} \\ \hline
Late Night      & 00:00 - 04:00  \\
Early Morning   & 04:00 - 08:00   \\
Morning         & 08:00 - 12:00  \\
Afternoon       & 12:00 - 16:00  \\
Evening         & 16:00 - 20:00  \\
Night           & 20:00 - 00:00  \\ \hline
\end{tabular}
% }
\end{table}

\begin{figure}[h!]
\centering
\includegraphics[width=0.7\columnwidth]{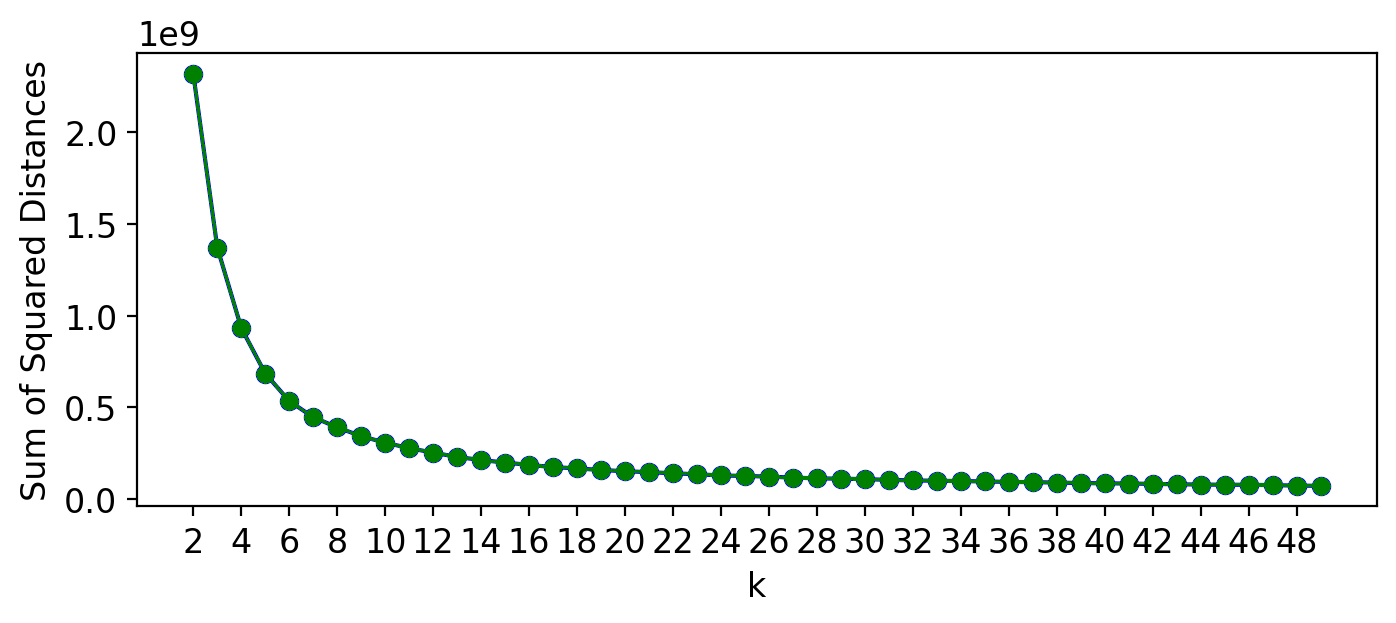}
\caption{Elbow method results}
\label{fig:elbow-method-results}
\end{figure}

We use the Elbow method that varies the number of clusters, \textit{k}, within a range to find the optimal number of clusters based on the sum of squared error \cite{syakur2018integration}. We vary \textit{k} from one to 50, as shown the Figure \ref{fig:elbow-method-results}. Based on the results, we selected $k=12$, because it is when the sum of squared error stabilizes.

\begin{figure}[h!]
\centering
\includegraphics[width=0.6\columnwidth]{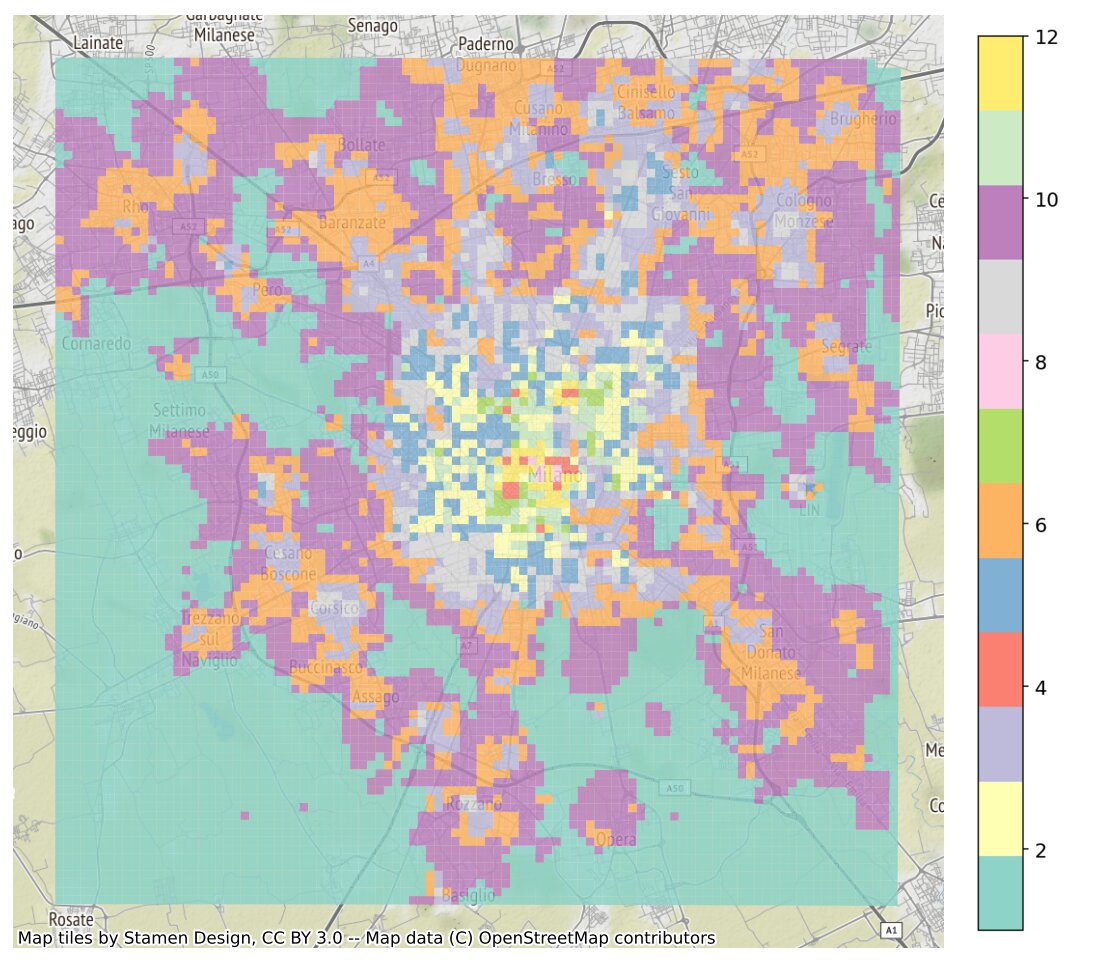}
\caption{Overlay of the 12 clusters on a map of the metropolitan area of Milan}
\label{fig:cluster}
\end{figure}

The Figure \ref{fig:cluster} shows the 12 clusters that have similar Internet activities overlaid in the Milan metropolitan area. One can note that the Cluster 1 represents more external areas and the Cluster 6 represents the center of the city of Milan (Figure \ref{fig:milan-region}). The other clusters represent different areas within the Milan region.

% We can use these clusters to train different agents using reinforcement learning algorithms. Each agent can learn the pattern related to a specific cluster in order to manage the SFC of EPC stack and handle the Internet activities.

We can train different agents using reinforcement learning algorithms to learn the best SFC allocation strategy for each cluster of cells, since different clusters have different Internet activities patterns. For example, the agent may allocate new VNFs when the traffic increases, and delete them (or some of them) when the traffic decreases, and as the cluster has different patterns, the the allocation strategy need be different as well.

\section{Reinforcement Learning for SFC Allocation}
\label{sec:rl-sfc}
The goal is to train a reinforcement learning agent to learn how to deploy a SFC taking into consideration the Internet activities using the data set described previously and the overall energy consumption. First of all, we need to define: the environment representation, the action representation, and the reward calculation.

The reinforcement learning agent needs to learn, for each state of the environment, what the action that will give the highest reward. This training consists of agent interactions with the environment over steps, where the agent takes actions that changes the environment state and receives rewards. For each step, we need to provide some relevant information about the environment for the agent. For a given state, the agent tries learn the best action that will return the highest reward. After the training, the agent learns the optimal policy, i.e., a set of actions for each state that will give the highest reward. For more information about the reinforcement learning theory, please see \cite{sutton2018reinforcement}.  

\subsection{Environment}
Our environment definition is composed of two main information: the Internet activities of all cells managed by the data centers, and the number of VNFs already allocated in the servers. To compose a step, we aggregated five minutes of the Internet activities from the Telecom Italia data set, because it is common to have periods when there is no Internet activity in several cells in the data set. We compose a vector with the number of Internet activities of all cells managed by the data centers. This information shows the number of Internet activities that will not be processed if the SFC is not running properly in the data centers, e.g., if there is no one or more VNFs that compose the SFC running. 

The other information given to the environment is the number of VNFs running in the servers, which is defined as a vector composed of the number of VNFs of each type present in each server of all data centers. As VNF types, we are considering the four components present in the EPC stack. This vector provides information about the current status of the servers with the purpose for choosing the best server to allocate a new VNF, for example. This vector can also provide information about the energy consumption of overall architecture, which is considered by the agent in the SFC allocation.

\subsection{Actions}
Given the environment definition, the agent needs to take actions, and each action has a specific reward for different states. The action representation is a vector of integers as defined in the Equation \ref{eq:action-representation}.

\begin{equation}
\label{eq:action-representation}
    Action = {(a_i,dc_i,s_i,vnf_i)}
\end{equation}

\noindent in which, $a_i$ represents the action type that the agent can take, and can be an integer, $a_i \in\{1,2,3,4\}$, where 1 is create a new VNF, 2 delete an existing VNF, 3 restart a VNF, and 4 does not change the environment; $dc_i$ and $s_i$ are, respectively, the data center and server ids where the VNF will be created, and need be defined by the agent; $vnf_i$ is the type of VNF, in our case, the four virtual functions (SGW, PGW, MME, and HSS). Therefore, given this representation, the agent can define actions of different types of VNF specifying the data center and its respective server. In cases where the agent selects the action 4, where the agent does not change the environment, the server and data center ids and the VNF type are not considered.

If the create action is selected, a new VNF will be created in the server for a given data center specified by the agent, since the this server have computational resources available. This resources limit is can be defined by a limited number of VNFs that can be allocated in a server, for example. If the restart action is selected, the probability failure of the VNF type selected is defined as zero and a new failure time is scheduled for this VNF.

It is important highlight that the agent just specifies the type of VNF, but for the restart and delete actions, the agent need to specify what VM will be affected (since there can have redundant VNFs of the same type). When one of these actions (restart or delete) is selected by the agent, we consider that the VNF with the highest fail probability will be affected (deleted or restarted), since our objective is to reduce the failures. For example, if in a given step, the agent takes the action with the specification $\{2,1,2,3\}$, the VNF of type 3 with the highest fail probability, in the server 2 of the DC 1 will be deleted.

\subsection{Reward}
After take the action, the agent receives a reward that will define the policy for taking new actions. The main objective of our agent is to reduce the number of VNF failures (and consequently the lost packets) considering the energy consumption of allocating new VNFs. The Equation \ref{eq:reward-equation} describes the reward calculation.

\begin{equation}
\label{eq:reward-equation}
    Reward=-((1-sfc_t)*packets_t)-\sum_{i=0}^{DCs}(energy_i) -restart_t + (sfc_t*f)
\end{equation}

in wich $sfc_t \in\{0,1\}$ represents the SFC status. If all VNFs that composes the SFC (PGW, SGW, MME, and HSS) are operational, then the SFC is complete and $sfc_t=1$. On the other hand, if there is no VNFs of at least one type running, the SFC is incomplete and $sfc_t=0$.
% in which $sfc_t \in\{1,0\}$ that represents if the whole SFC is not operational at step $t$, i.e., if one or more out of the four VNFs (PGW,SGW, MME, or HSS) is not operational. 
$sfc_t$ is equals to 0 at the beginning of the system operation, where there is no VNFs allocated, and also if the agent deletes all VNFs of a specific type in all data centers. In this case, the agent receives a negative reward multiplied by the number of Internet activities, $packets_t$, of all cells for a given step $t$ in order to increase the penalty according with the network status. We also apply a penalty for the agent according to the energy consumption of the VFNs allocated. We consider the energy consumption of data center $i$, $energy_i$, as defined in the Equation \ref{eq:energy}, where $CPU_k$ and $MEM_k$ are the energy consumption of a CPU and a memory, respectively, in terms of allocate a new VNF at server $j$. We also apply a little penalty if the agent decide to restart a VNF defined by the variable $restart_t \in \{0,1\}$, with the purpose not stimulate the agent restart the VNFs frequently.

\begin{equation}
\label{eq:energy}
    energy_i=\sum_{j=0}^{servers}\sum_{k=0}^{vnfs}CPU_k+MEM_k
\end{equation}

Finally, the agent receives a positive reward if the SFC is complete ($sfc_t=1$), that is, if all VNFs that defines a SFC are running in the servers of data centers.
%Finally, the agent receives a positive reward if the SFC is complete (represented by ($1-sfc_t$)), since the agent create all VNFs of the SFC and whether no server or VNF failure impacted the operation of the SFC as a whole. 
The factor $f$ (we consider equals to 100) is applied to adjust the reward for the agent, where the greater the factor, the greater the agent's reward if the SFC is operational.

\section{Preliminary Results}
\label{sec:results}
To create the environment described previously, we used the framework Stable Baselines\footnote{\url{https://stable-baselines.readthedocs.io/en/master/index.html}}. The Stable Baselines is a fork of OpenAI baselines project\footnote{\url{https://github.com/openai/baselines}} with new algorithms, wrappers for preprocessing and multiprocessing, and friendly interfaces to create new agents and environments. 

To create the agent, we used the standard implementation of PPO algorithm provided by the Stable Baselines framework. The PPO combines ideas from Advantage Actor Critic (A2C) \cite{mnih2016asynchronous} and Trust Region Policy Optimization (TRPO) \cite{schulman2015trust} algorithms and outperformed them in complex tasks such as robotic locomotion and Atari game playing. For more details about PPO, see (\cite{schulman2017proximal}).

We implemented a simulator using Python 3.6 language to simulate the events of incoming Internet activities (based on the data set described in the Subsection \ref{subsed:tim-dataset}) and the failure and repair events of the servers and VNFs. This simulator is used to implement the steps of the environment. At begin of the simulation, the servers are created on the data centers and a failure time is scheduled to happen for each server. When a VNF is created by the agent, a failure time is scheduled to happen similar to the server. After a failure event, this component becomes unavailable and a repair event of the component is scheduled. After the repair event, the component becomes available again and a failure event is scheduled. The failure and repair events are generated following exponential distributions \cite{andrade2017availability}, which are defined by mean time to failure (MTTF) and mean time to repair (MTTR) values, respectively.

We consider a simple scenario to evaluate the PPO performance to allocate SFC in the data centers in order to reduce the number of failure and lost packets. Table \ref{tab:scenario-configuration} summarizes the scenario configuration. 

\begin{table}[h]
\caption{Scenario Configuration}
\label{tab:scenario-configuration}
\centering
\begin{tabular}{@{}cc@{}}
\toprule
\textbf{Parameter}                                                          & \textbf{Value} \\ \midrule
Number of data centers                                                      & 10             \\
\midrule
\begin{tabular}[c]{@{}c@{}}Number of servers\\ per data center\end{tabular} & 5              \\
\midrule
\begin{tabular}[c]{@{}c@{}}Maximum number \\ of redundant VNFs\end{tabular} & 5              \\
\midrule
MTTF server                                                               & 8760 hours \cite{santos2018analyzing}     \\
\midrule
MTTR server                                                                 & 1.667 hours \cite{santos2018analyzing}    \\
\midrule
MTTF VNF                                                                    & 24 hours \cite{araujo2016software}      \\
\midrule
MTTR VNF                                                                    & 0.033 hours    \\
\midrule
CPU energy consumption                                                      & 40 W \cite{ali2015energy}          \\
\midrule
Memory energy consumption                                                   & 30.72 W \cite{ali2015energy}        \\ \bottomrule
\end{tabular}
\end{table}

We consider the cells of Cluster 5 (see Figure \ref{fig:cluster}) which corresponds to the center of Milan region, and has the highest Internet activity volume. This cluster is composed of 276 cells. We consider 10 data centers to manager these cells, each one with five servers dedicated to the SFC of EPC stack. We distributed them equally through the cells. Due to limited resources, we assume that each server can have a maximum of five VNFs allocated. We also assume that there is a limit of redundant VNFs allocated of the same type in one server, in order to increase the heterogeneity of VNFs in the same server. We consider the energy consumption of CPU and memory in Watts, as show in the Table \ref{tab:scenario-configuration}.

Due to the stochastic characteristics of our simulation and the agent training process, we carried out 100 experiments, and the results presented in this paper are the average.
We defined the number of steps according with the Telecom Italia data set events. As we aggregated the Internet activities for each five minutes and the data set has data about 62 days, we have 8,928 events. We divided these events in 90\% for create an environment to train the agent (8,035 steps), and 10\% to test the agent (893 steps).

\begin{figure}[h!]
\centering
\includegraphics[width=0.6\columnwidth]{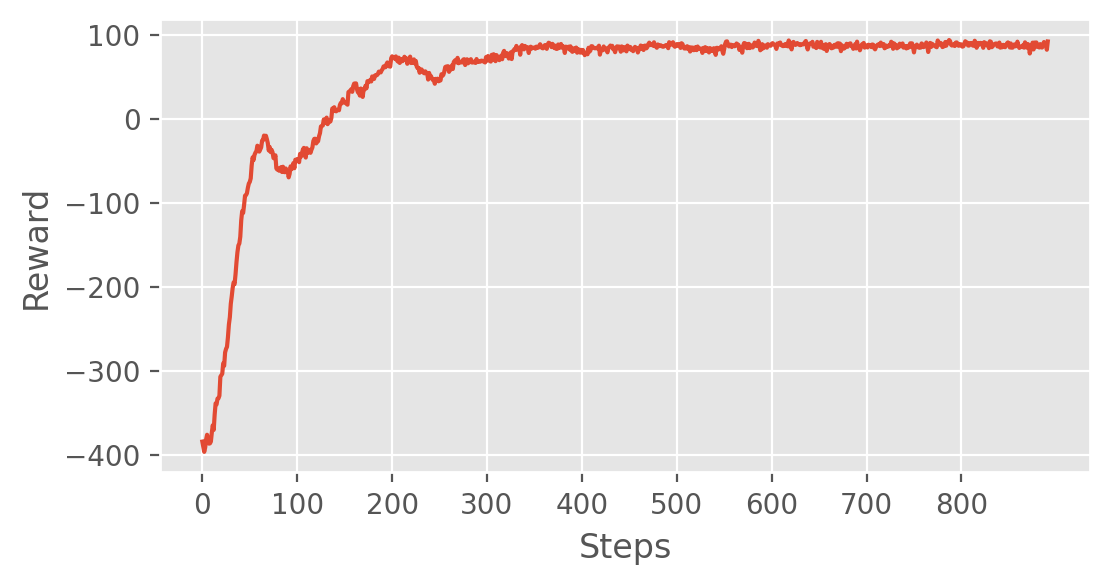}
\caption{Reward achieved by agent over time}
\label{fig:reward-results}
\end{figure}

Figure \ref{fig:reward-results} shows the cumulative reward achieved by the agent over the steps. One can note that, at the first step, the reward is around -400, and it is justified because at the beginning of the simulation, there is not VNFs allocated yet, then the SFC is incomplete, losing all packets from the cells. But the agent learns quickly that this situation is not suitable and creates VNFs to allocate the whole SFC in the data centers. The agent is able to allocate the entire SFC around step 80, when the reward approaches zero. After that, the reward drops a little, this may be justified by the energy consumption of the allocated NFVs, but the reward starts to grow again soon afterwards. After step 200, the reward stabilizes and remains close to 100 (value of factor $f$ defined in the Equation \ref{eq:reward-equation} when the SFC is complete). One can note that the reward stabilizes and has slight osculations (with a little decreasing around the step 250), it happens due to the allocation of new VNFs, that increases the energy consumption and, consequently, receives negative rewards. These oscillations can be caused by little SFC unavailability, due to eventual failures in the servers and VNFs, but the impact is low.

%Figure \ref{fig:reward-results} shows the reward achieved by the agent over the steps. One can note that, at the first step, the reward is around -400, and it is justified because at the beginning of the simulation, there is not VNFs allocated yet, then SFC is incomplete and resulting in lost all packets from the cells. But the agent learns quickly that this situation is not suitable and creates VNFs to allocate the whole SFC in the data centers. The agent is able to allocate the entire SFC around step 80, when the reward becomes positive (with a little decrease afterwards). After step 200, the reward stabilizes and remains close to 100 (value of factor $f$ defined in the Equation \ref{eq:reward-equation} when the SFC is complete). One can note that the reward stabilizes and has slight osculations, it happens due to the allocation of new VNFs, that increases the energy consumption and, consequently, receives negative rewards. These oscillations can be caused by little SFC unavailability, due to eventual failures in the servers and VNFs, but the impact is low.

\begin{figure}[h!]
\centering
\includegraphics[width=0.6\columnwidth]{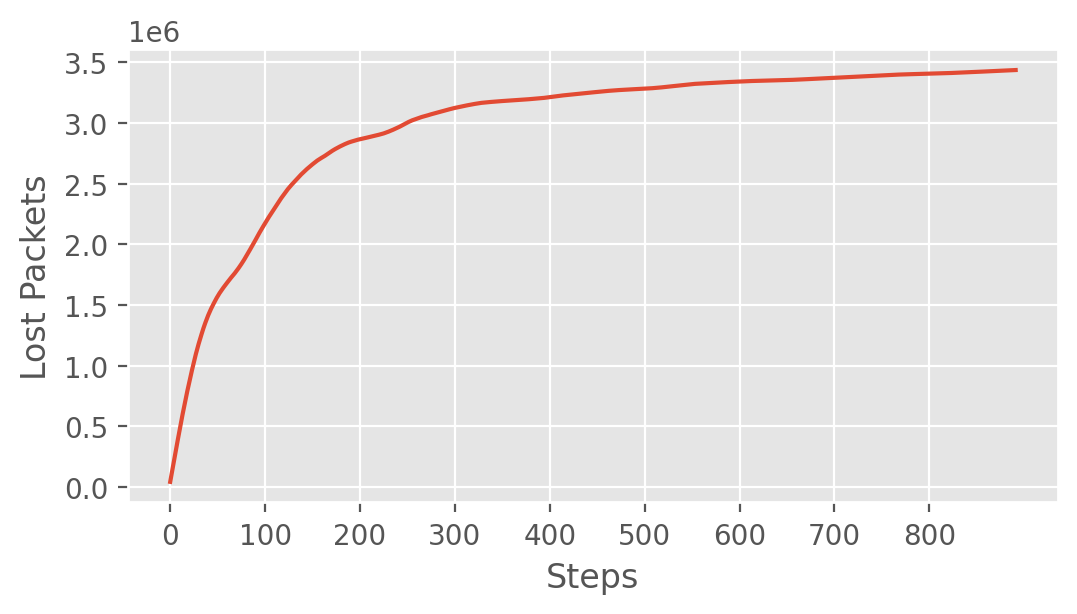}
\caption{Accumulated lost packets}
\label{fig:lost-packets-results}
\end{figure}

Figure \ref{fig:lost-packets-results} shows the accumulated lost packets over the steps. Up to step 100, it can be seen that the growth is very fast, given that all packages are being lost, as the SFC is not fully allocated yet. After step 200, the number of packets lost starts to grow slowly. This stabilization can be attributed to the SFC management allocated by the agent, since it can restart the VNFs that are about to fail.

\section{Conclusion}
\label{sec:conclusion}

In this paper, we presented a solution to automatically allocate and manage SFC in a cellular network context. The main goal was to allocate SFC and manage VNFs that may fail in order to reduce the number of failures and, consequently, the number of lost packets. The solution takes into account the energy consumption as constraints. We represented a cellular network scenario based on real data set. We also created a simulator to represent the failure and repair events of servers and the VNFs allocated over distributed data centers. We solved the SFC allocation and the VNF management problem using a reinforcement learning algorithm.

We carried out simulations using the PPO implementation provided by the Stable Baselines algorithms, and preliminary results showed that the PPO is able to allocate a complete SFC after some steps and manage the VNFs controlling the lost packets taking into account the energy consumption.

We have several future works to improve the solution presented in this paper. We plan to consider a scenario where many SFCs have to be allocated and managed concurrency. This feature increases the scenario complexity, since SFCs need to be allocated in limited computational resources. We also plan to consider the bandwidth of link between the servers in the VNF placement. New architectures can be considered, where groups of data centers can be considered to manage different regions of a city, for example. Finally, we plan to compare other reinforcement learning algorithms and make a parameter variation to evaluate different configurations.

\bibliographystyle{unsrt}  

\bibliography{references}

\end{document}